\documentclass[times,11pt]{article}

\usepackage{fullpage,amsmath,amssymb,amsthm}

\newcommand{\Fi}{\mathbb{F}_p}
\newtheorem{theorem}{Theorem}
\newtheorem{definition}{Definition}
\newtheorem{lemma}{Lemma}
\newcommand{\cH}{\mathcal{H}}
\newcommand{\eps}{\varepsilon}

\DeclareMathOperator{\poly}{poly}
\renewcommand{\log}{\lg}

\begin{document}

\title{Optimal Non-Adaptive Cell Probe Dictionaries and Hashing\footnote{This paper is a merge and revision of two previous reports~\cite{PersianoYao} and~\cite{larsen2023optimal}.}}

\author{
Kasper Green Larsen\\Aarhus University \\ \textsf{larsen@cs.au.dk}
\and
Rasmus Pagh\\University of Copenhagen \\ \textsf{pagh@di.ku.dk} 
\and
Giuseppe Persiano\\Universit\`a di Salerno\\ \textsf{giuper@gmail.com}
\and
Toniann Pitassi\\Columbia University\\ \textsf{tonipitassi@gmail.com}
\and
Kevin Yeo\\Columbia University\\ \textsf{kwlyeo@cs.columbia.edu}
\and 
Or Zamir\\Tel Aviv University \\ \textsf{orzamir@tauex.tau.ac.il}
}

\date{}

\maketitle

\begin{abstract}
    We present a simple and provably optimal non-adaptive cell probe data structure for the static dictionary problem. Our data structure supports storing a set of $n$ key-value pairs from $[u]\times [u]$ using $s$ words of space and answering key lookup queries in $t = O(\lg(u/n)/\lg(s/n))$ non-adaptive probes. This generalizes a solution to the membership problem (i.e., where no values are associated with keys) due to Buhrman et al. We also present matching lower bounds for the non-adaptive static membership problem in the deterministic setting. Our lower bound implies that both our dictionary algorithm and the preceding membership algorithm are optimal, and in particular that there is an inherent complexity gap in these problems between no adaptivity and one round of adaptivity (with which hashing-based algorithms solve these problems in constant time).

    Using the ideas underlying our data structure, we also obtain the first implementation of a $n$-wise independent family of hash functions with optimal evaluation time in the cell probe model. 
\end{abstract}

\newpage

\section{Introduction}
\label{sec:intro}
The static membership problem is arguably the simplest and most fundamental data structure problem. In this problem, the input is a set $S$ of $n$ integer keys $x_1,\dots,x_n \in [u] = \{0,\dots,u-1\}$ and the goal is to store them in a data structure, such that given a query key $x \in [u]$, the data structure supports reporting whether $x \in S$.

The classic solution to the membership problem is to use \emph{hashing}, suggested as early as by Tarjan-Yao~\cite{tarjan1979storing}. 
The textbook hashing-based solution is hashing with chaining, where one draws a random \emph{hash function} $h : [u] \to [m]$ and creates an array $A$ with $m = O(n)$ entries. Each entry $A[i]$ of the array stores a linked list of all keys $x \in S$ such that $h(x) = i$. To answer a membership query for $x$, we compute $h(x)$ and scan the linked list in entry $A[h(x)]$. If $h$ is drawn from a \emph{universal} family of hash functions, the time to answer queries is $O(1)$ in expectation. 
The expected query time can be made worst case $O(1)$ using e.g.~perfect hashing~\cite{FredmanKS84} or (static) Cuckoo hashing~\cite{Pagh01,PaghR01}. 
All of the above solutions may also be easily extended to solve the dictionary problem in which the data to be stored is a set of $n$ key-value pairs $\{(x_i,y_i)\}_{i=1}^n$. Upon a query $x$, the data structure must return the value $y_i$ such that $x_i = x$, or report that no such pair exists.  

\subsection{Adaptivity and Membership}
A common feature of all hashing based solutions to the membership and dictionary problem, is that they are \emph{adaptive}. That is, the memory locations they access depend heavily on the random choice of hash functions.
In particular, to answer a query we first need to read the description of the chosen hash function, and only based on that we can compute the next memory cells we should access.
A \emph{non-adaptive} data structure has the property that the memory cells to access on a query $x$ are completely determined from $x$ itself. 
Non-adaptive data structures are studied for several reasons, a common type being computational settings in which interaction with memory is either expensive or limited.
Non-adaptive data structures allow retrieving all necessary memory cells in \emph{parallel} when answering a query, circumventing any memory-access related latency.
This property also allows simpler implementation of the data structure under cryptographic settings, such as encrypted computation with Fully Homomorphic Encryption (see~\cite{cryptocuckoo} for more details on the importance
of non-adaptive querying in cryptography).

In this work, we present a non-adaptive dictionary algorithm in which a query needs to only access logarithmically many memory cells, and also prove a matching lower bound (which holds even for the static membership problem).

Unlike the textbook solution of hashing with chaining, which requires many rounds of adaptivity due to scanning a linked list, other solutions (e.g., cuckoo hashing) only need one round of adaptivity (i.e., first they read the description of the hash function, and then read memory cells that are determined only by the query and the hash function).
Our results imply that a single round of adaptivity is necessary and sufficient to reduce the query time from super-constant to constant.

\medskip\noindent{\bf The Cell Probe Model.}
The cell probe model by Yao~\cite{Yao81a} is the de-facto model for proving data structure lower bounds. In this model, a data structure consists of a memory of $s$ cells with integer addresses $0,\dots,s-1$, each storing $w$ bits. Computation is free of charge in this model and only the number of memory cells accessed/probed when answering a query counts towards the query time. A lower bound in the cell probe model thus applies to any data structure implementable in the classic word-RAM upper bound model.

\medskip\noindent{\bf Previous Work.}
Buhrman, Miltersen, Radhakrishnan and Venkatesh~\cite{Buhrman02} showed that it is possible to store a data structure of size $O(n\log u)$ bits such that membership queries can be answered in $O(\log u)$ non-adaptive \emph{bit} probes (i.e., the cell probe model with $w=1$).
This of course implies a membership data structure with $O(\log u)$ probes in the cell probe model, but it is not clear how to extend it to solve the dictionary problem with the same time and space complexity. Furthermore, the data structure by Buhrman et al. is non-explicit in the sense that they give a randomized argument showing existence of an efficient data structure.
Buhrman et al.~also show a lower bound of $t = \Omega(\lg(u/n)/\lg(s/n))$ bit probes. In the setting where $n$ is polynomially smaller than $u$ and $s$ is $O(n)$ this matches the upper bound up to constant factors (but it is possible that a tighter analysis can be made).
Alon and Feige~\cite{alon2009power}
as well as Garg and Radhakrishnan~\cite{garg2017set}
studied space lower bounds for dictionary data structures with three non-adaptive probes in the bit probe model. The best lower bound shows that space of $s = \Omega(\sqrt{u n})$ is necessary. 


Berger et al.~\cite{Berger2006} study the non-adaptive dictionary problem, but in the I/O model, i.e., a single memory access can retrieve $B\geq 1$ keys or values.
In the Word RAM model this corresponds to having word size $B\log u$.
This means that their strongest results for the dictionary problem would require word size $\Omega(\log(n)\log(u))$ --- as we will see later, our results hold for word size $\log u$.

Brody et al.~\cite{BrodyLarson} present a \emph{dynamic} non-adaptive data structure for the \emph{predecessor search} problem, allowing insertions and deletions of keys while supporting predecessor queries in $O(\lg u)$ probes. A predecessor query for a key $x$ must return the largest $x' \in S$ such that $x' \leq x$. Such a data structure clearly also supports membership queries. However, their data structure critically uses $s = \Theta(2^w) = \Theta(u)$ memory. For the membership problem in this setting, a bit-vector with constant time operations suffices.
Brody et al.~\cite{BrodyLarson} however prove that for dynamic data structures
for predecessor search, this query time is optimal even with $\Theta(u)$ space.
Boninger et al.~\cite{BBK17}
as well as Ramamoorthy and Rao~\cite{RR18} also study lower bounds
for the non-adaptive dynamic predecessor problem.
Relating their results to the non-adaptive static dictionary problem, the two works show
query time must be $t = \Omega(\log u / \log w)$ and $t = \Omega(\log u / (\log\log u + \log w))$ respectively in the cell probe model. To our knowledge, these are the highest known lower bounds for the static, non-adaptive dictionary problem.

This still leaves open the problem of obtaining an optimal static and non-adaptive membership
data structure, in both the word-RAM model, and in the cell probe model.


\medskip\noindent{\bf Our Contribution.}
In this work, we present a simple and optimal non-adaptive cell probe data structure for the dictionary and membership problem:
\begin{theorem}
\label{thm:dic}
For any $s = \Omega(n)$, there is a non-adaptive static cell probe data structure for the dictionary problem, storing $n$ key-value pairs $(x_i,y_i) \in [u] \times [u]$ using $s$ memory cells of $w = \Theta(\lg u)$ bits and answering queries in $t = O(\lg(u/n)/\lg(s/n))$ probes.
\end{theorem}

As stated in the theorem, our data structure is implemented in the cell probe model, meaning that we treat computation as free of charge. Implementing the data structure in the more standard upper bound model, the word-RAM, would require the construction of a certain type of explicit bipartite expander graph.

Compared to prior works (such as~\cite{Buhrman02,Berger2006}), our construction shows that we may rely on a significantly weaker expansion argument. Past constructions required an orientability argument to assign
memory to keys that required expanders with a strong unique-neighbors property. In contrast, our construction utilizes weaker non-contractive expanders
to argue that there is sufficient capacity to accommodate storage of all keys (using Hall's theorem). This directly translates to a logarithmic improvement in space usage.
Namely, we only require the existence of $t$-left-regular bipartite graphs with expansion factor one; however our bipartite graph is highly imbalanced. 
Our expansion property corresponds to an imbalanced disperser, and therefore is well-studied and has other applications (e.g., \cite{GuruswamiUV09}). 
Such dispersers exist by a counting argument, but it remains an open problem to obtain explicit constructions. 
A recent work~\cite{bauwens2022hall} constructs explicit expanders that may
be plugged into our construction to obtain an explicit RAM upper bound. However, this incurs a poly-logarithmic blowup and obtaining a tight explicit
RAM upper bound would require better explicit expanders.


We also present a matching lower bound for the non-adaptive dictionary and membership problem in the cell probe model:

\begin{theorem}
\label{thm:det_lb}
For any non-adaptive static cell probe data structure for the dictionary problem storing $n$ key-value pairs $(x_i, y_i) \in [u] \times [u]$ using $s$ memory cells of $w$ bits and answering queries in $t$ probes must satisfy
$$
    t = \Omega\left(\min\left\{ \frac{ n \lg(u/n)}{w}, \frac{\lg(u/n)}{\lg(sw/(n \lg(u/n)))}\right\}\right).
$$
\end{theorem}

Our lower bound shows that adaptivity is crucial to obtain constant query time. In particular, non-adaptive data structures require super-constant query time while well-known constructions with adaptivity (such as cuckoo hashing) can obtain constant query time.

We note that our lower bound peaks higher compared to the prior best lower bounds. For standard parameters of $u = n^{1 + O(1)}$ and $w = \Theta(\log u)$, our lower bound shows that optimal space constructions with $s = O(n)$ require query time $t = \Omega(\log u)$ in the cell probe model. In contrast, prior works~\cite{BBK17,RR18} obtain
lower bounds of $t = \Omega(\log u / \log\log u)$.

\subsection{Hash Functions with High Independence}
When using hash functions in the design of data structures and algorithms, it is often assumed for simplicity of analysis that truly random hash functions are available. Such a hash function $h : [u] \to [m]$ maps each key \emph{independently} to a uniform random value in $[m]$. Or said differently, when drawing the random hash function $h$, we choose a uniform random function in the family of hash functions $\cH$ consisting of all (deterministic) functions from $[u]$ to $[m]$. Implementing such a hash function in practice is often infeasible as it requires $u \lg m$ random bits and thus the storage requirement may completely dominate that of any data structure making use of the hash function.

Fortunately, much weaker hash functions suffice in many applications. The simplest property of a family of hash functions $\cH \subseteq [u] \to [m]$, is that it is \emph{universal}~\cite{carter1977universal}. A universal family of hash functions has the property that for a uniform random $h \in \cH$, it holds for every pair of keys $x \neq y \in [u]$ that $\Pr[h(x)=h(y)] \leq 1/m$. Universal hashing for instance suffices for implementing hashing with chaining with expected constant time membership queries, but is not sufficient for implementing Cuckoo hashing~\cite{CohenKane09}. The next step up from universal hashing is the notion of $n$-wise independent hashing. A family of hash functions $\cH$ is $n$-wise independent if, for $h$ drawn uniformly from $\cH$, it holds for any set of $n$ distinct keys $x_1,\dots,x_n$ that $h(x_1),\dots,h(x_n)$ are independent and uniformly random (or nearly uniformly random). The prototypical example of an $n$-wise independent family of hash function (with nearly uniform hash values) is 
$$
\cH := \left\{h_{\alpha_0,\dots,\alpha_{n-1}}(x) = \left(\sum_{i=0}^{n-1} \alpha_i x^i \bmod p\right) \bmod m \mid \alpha_0,\dots,\alpha_{n-1} \in [p] \right\}
$$
where $p$ is any prime greater than or equal to $u$. That is, to draw a hash function $h$ from $\cH$, we sample $\alpha_0,\dots,\alpha_{n-1}$ uniformly and independently in $[p]$ and let $h(x)$ be the evaluation of the polynomial $(\sum_i \alpha_i x^i \bmod p) \bmod m$.\footnote{Technically, this hash function is only approximately $n$-wise independent, in the sense that the hash values of any $n$ keys \emph{are} independent, but only approximately uniform random.} Clearly, the evaluation time of this hash function is $\Theta(n)$. Whether it is possible to implement $n$-wise independent hash functions with faster evaluation time has been the focus of much research. On the lower bound side, Siegel~\cite{Siegel89} proved that any implementation of an $n$-wise independent hash function $h : [u] \to [m]$ using $s$ memory cells of $w = \Theta(\lg u)$ bits, must probe at least $t = \Omega(\min\{\lg(u/n) /\lg(s/n),n\})$ memory cells to evaluate $h$. The hash function above matches the second term in the minimum. For the first term, the result that comes closest is a recursive form of tabulation hashing by Christiani et al.~\cite{ChristianiPT15} that gives an $n$-wise independent family of hash functions that can be implemented using $s=O(n u^{1/c})$ space and evaluation time $t=O(c \lg c)$ for any $c = O(\lg u/\lg n)$. Rewriting the space bound gives $c = \lg u/\lg(s/n)$ and thus $t = O(\lg(u) \lg(\lg(u)/\lg(s/n))/\lg(s/n))$. This is about a $\lg \lg u$ factor away from the lower bound of Siegel in terms of the query time $t$. This algorithm is adaptive and requires $s \geq n^{1+\Omega(1)}$ as they need $\lg u/\lg(s/n) = O(\lg u/\lg n)$. 

\medskip\noindent{\bf Our Contribution.}
Designing an optimal $n$-wise independent family of hash functions thus remains open, with or without adaptivity. In this work, we show how to implement such a function in the cell probe model (where computation is free):
\begin{theorem}
\label{thm:hash}
    For any $s = \Omega(n)$ and $p = \Omega(u)$, there is a non-adaptive static cell probe data structure for storing an $n$-wise independent hash function $h : [u] \to \Fi$ using $s$ memory cells of $w = \Theta(\lg p)$ bits and answering evaluation queries in $t = O(\lg(u/n)/\lg(s/n))$ probes.
\end{theorem}
We remark that Siegel's lower bound holds in the cell probe model, and thus our data structure is optimal. Furthermore, Siegel's lower bound holds \emph{also} for adaptive data structures, whereas ours is even non-adaptive. Compared to the work of Christiani et al., we have a faster evaluation time and only require $s = \Omega(n)$. The downside is of course that our solution is only implemented in the cell probe model. Implementing our hash function in the word-RAM model would require the same type of explicit expander graph as for implementing our non-adaptive dictionary (and a bit more), further motivating the study of such expanders (see Section~\ref{sec:concl}).

To compare with previous techniques, we note that the majority
of prior works (such as~\cite{pagh2003uniform,dietzfelbinger2003almost,ChristianiPT15}) consider
adaptive constructions.
The original work of Siegel~\cite{Siegel89} did not directly study
non-adaptivity. However, Lemma 2 in~\cite{Siegel89} can be used to
construct a non-adaptive construction in the cell probe model using
a suitable expander graph. Our construction leads to a better (and tight)
upper bound in addition to being simpler by replacing polynomials with a simple sum of memory cells.

\section{Non-Adaptive Dictionaries}
We consider the dictionary problem where we are to preprocess a set $X$ of $n$ key-value pairs from $[u] \times [u]$ into a data structure, such that given an $x \in [u]$, we can quickly return the corresponding value $y$ such that $(x,y) \in X$ or conclude that no such $y$ exists. We assume that any for any key $x$, there is at most one value $y$ such that $(x,y) \in X$.

We focus on non-adaptive data structures in the cell probe model. Non-adaptive means that the memory cells probed on a query depends only on $x$. We assume $u = \Omega(n)$ and that the cell size $w$ is $\Theta(\lg u)$.

As mentioned in Section~\ref{sec:intro}, we base our data structure on expander graphs. We recall the standard definitions of bipartite expanders in the following:

\begin{definition}
    A $(u,s,t)$-bipartite graph with $u$ left vertices, $s$ right vertices and left degree $t$ is specified by a function $\Gamma : [u] \times [t] \to [s]$, where $\Gamma(x,y)$ denotes the $y^{th}$ neighbor of $x$. For a set $S \subseteq [u]$, we write $\Gamma(S)$ to denote its neighbors $\{\Gamma(x,y) : x \in S, y \in [t]\}$.
\end{definition}

\begin{definition}
    A bipartite graph $\Gamma : [u] \times [t] \to [s]$ is a $(K,A)$-expander if for every set $S \subseteq [u]$ with $|S|=K$, we have $|\Gamma(S)| \geq A \cdot K$. It is a $(\leq K_{\max},A)$-expander if it is a $(K,A)$-expander for every $K \leq K_{\max}$.
\end{definition}

The literature on bipartite expanders, see e.g.~\cite{GuruswamiUV09}, is focused on graphs with near-optimal expansion $A = (1-\eps)t$, i.e. very close to the largest possible expansion with degree $t$. However, for our non-adaptive dictionaries, we need significantly less expansion. We call such expanders \emph{non-contractive} and define them as follows:

\begin{definition}
    A bipartite graph $\Gamma : [u] \times [t] \to [s]$ is a $(\leq K_{\max})$-non-contractive expander if it is a $(\leq K_{\max}, 1)$-expander.
\end{definition}

Said in words, a bipartite is a $(\leq K_{\max})$-non-contractive expander, if every set of at most $K \leq K_{\max}$ left-nodes has at least $K$ neighbors.

Before presenting our dictionary, we present the second ingredient in our dictionary, namely Hall's marriage theorem. For a bipartite graph with left-vertices $X$, right-vertices $Y$ and edges $E$, an $X$-perfect matching is a subset of disjoint edges from $E$ such that every vertex in $X$ has an edge. Hall's theorem then gives the following:
\begin{theorem}[Hall's Marriage Theorem]\label{thm:hall}
    A bipartite graph with left-vertices $X$ and right-vertices $Y$ has an $X$-perfect matching if and only if for every subset $S \subseteq X$, the set of neighbors $\Gamma(S)$ satisfies $|\Gamma(S)| \geq |S|$.
\end{theorem}

With these ingredients, we are ready to present our dictionary.

\medskip\noindent{\bf Dictionary from Non-Contractive Expander.}
Given a set of $n$ key-value pairs $X = \{(x_i,y_i)\}_{i=1}^n \subset [u] \times [u]$ and a space budget of $s$ memory cells, we build a data structure as follows:

\textit{Construction.}
Initialize $s$ memory cells and let $\Gamma : [u] \times [t] \to [s]$ be a $(\leq n)$-non-contractive expander for some $t$. Construct the bipartite graph $G$ with a left-vertex for each $x_i$ and a right vertex for each of the $s$ memory cells. Add an edge from $x_i$ to each of the nodes $\Gamma(x_i,j)$ for $i=0,\dots,t-1$. Note that this is a subgraph of the bipartite $(\leq n)$-non-contractive expander corresponding to $\Gamma$. It follows that for every subset $S \subseteq \{x_i\}_{i=1}^n$, we have $|\Gamma(S)| \geq |S|$. We now invoke Hall's Marriage Theorem (Theorem~\ref{thm:hall}) to conclude the existence of an $\{x_i\}_{i=1}^n$-perfect matching on $G$. Let $M = \{(x_i, v_i)\}_{i=1}^n$ denote the edges of the matching. For each such edge $(x_i,v_i)$, we store the key-value pair $(x_i,y_i)$ in the memory cell of address $v_i$. For all remaining $s-n$ memory cells, we store a special \textit{Nil} value.

\textit{Querying.} Given a query $x \in [u]$, we query the $t$ memory cells of address $\Gamma(x,i)$ for $i=0,\dots,t-1$. If any of them stores a pair $(x,y)$, we return $y$. Otherwise, we return \textit{Nil} to indicate that no pair $(x,y)$ exists in $X$. 

\textit{Analysis.} Correctness follows immediately from Hall's Marriage Theorem. The space usage is $s$ memory cells of $w = \Theta(\lg u)$ bits and the query time is $t$. The required perfect matching $M$ can be computed in $\poly(n,s)$ times after performing $O(nt)$ queries to obtain the edges of the subgraph induced by the left-vertices $\{x_i\}_{i=1}^n$. We thus have the following result:
\begin{lemma}
\label{lem:dicfromexp}
    Given a bipartite $(\leq n)$-non-contractive expander $\Gamma : [u] \times [t] \to [s]$, there is a non-adaptive dictionary for storing a set of $n$ key-value pairs using $s$ cells of $w=\Theta(\lg u)$ bits and answering queries in $t$ evaluations of $\Gamma$ and $t$ memory probes. The dictionary can be constructed in $\poly(n,s)$ time plus $O(nt)$ evaluations of $\Gamma$.
\end{lemma}

Lemma~\ref{lem:dicfromexp} thus gives us a way of obtaining a non-adaptive dictionary from an expander. What remains is to give expanders with good parameters. As mentioned, we do not have optimal explicit constructions of such expanders. However, for the cell probe model where computation is free of charge, we merely need the existence of $\Gamma$ and not that it is efficiently computable. Concretely, a probabilistic argument gives the following:
\begin{lemma}
\label{lem:noncons}
    For any $s \geq 2n$ and any $u \geq n$, there exists a (non-explicit) $(\leq n)$-non-contractive expander $\Gamma : [u] \times [t] \to [s]$ with $t = \lg(u/n)/\lg(s/n) + 5$.
\end{lemma}
Combining Lemma~\ref{lem:dicfromexp} and Lemma~\ref{lem:noncons} implies our Theorem~\ref{thm:dic}.

\medskip\noindent{\bf Non-Explicit Expander.}
In the following, we prove Lemma~\ref{lem:noncons}. For this, consider drawing $\Gamma : [u] \times [t] \to [s]$ uniformly among all such functions/expanders. That is, we let $\Gamma(x,y)$ be uniform random and independently chosen in $[s]$ for each $x \in [u]$ and $y \in [t]$. For each $S \subseteq [u]$ with $|S|\leq n$ and each $T \subseteq [s]$ with $|T|=|S|-1$, define an event $E_{S,T}$ that occurs if $\Gamma(S) \subseteq T$. We have that $\Gamma$ is a $(\leq n)$-non-contractive expander if none of the events $E_{S,T}$ occur. For a fixed $E_{S,T}$, we have $\Pr[E_{S,T}] = (|T|/s)^{t|S|}$ and thus a union bound implies
\begin{eqnarray*}
    \Pr[\Gamma \textrm{ is not a $(\leq n)$-non-contractive expander}] &\leq& \\
    \sum_{S,T} \Pr[E_{S,T}] 
    &=& \\
    \sum_{i=1}^n \sum_{S \subseteq [u] : |S|=i} \sum_{T \subseteq [s] : |T|=i-1} \Pr[E_{S,T}] 
    &\leq& \\
    \sum_{i=1}^n \binom{u}{i} \binom{s}{i} (i/s)^{t i} 
    &\leq& \\
    \sum_{i=1}^n (eu/i)^i (es/i)^i (i/s)^{ti} &=& \\
    \sum_{i=1}^n \left(\frac{e^2 u i^{t-2}}{s^{t-1}}\right)^i &\leq& \\
    \sum_{i=1}^n \left(e^2 (u/n) (n/s)^{t-1}\right)^i.
\end{eqnarray*}
For $s \geq 2n$ and $t \geq \lg(u/n)/\lg(s/n)+5$, this is at most $\sum_{i=1}^n (e^2/16)^i < 1$ and thus proves Lemma~\ref{lem:noncons}.

\section{Hashing}
\label{sec:hash}
In this section, we show how to construct a $n$-wise independent hash function with fast evaluation in the cell probe model. As a data structure problem, such a data structure has a query $h(x)$ for each $x \in [u]$. Upon construction, the data structure draws a random seed and initializes $s$ memory cells of $w$ bits. The data structure satisfies that the values $h(x)$ are uniform random in $\Fi$ and $n$-wise independent. Here the randomness is over the choice of random seed.

Similarly to our dictionary, our hashing data structures makes use of a bipartite expander. However, we need a (very) slightly stronger expansion property. Concretely, we assume the availability of a $(\leq n, 2)$-expander $\Gamma : [u] \times [t] \to [s]$ (rather than a $(\leq n,1)$-expander). The expander $\Gamma$ thus satisfies that for any $S \subseteq [u]$ with $|S| \leq n$, we have $|\Gamma(S)| \geq 2|S|$. 

In addition to the $(\leq n,2)$-expander $\Gamma$, we also need another function assigning \emph{weights} to the edges of $\Gamma$. We say that $\Pi : [u] \times [t] \to \Fi$ makes $\Gamma$ \emph{useful} if the following holds: Construct from $(\Gamma,\Pi)$ the $u \times s$ matrix $A_{\Gamma,\Pi}$ such that entry $(x,y)$ equals
\[
\sum_{j : \Gamma(x,j)=y} \Pi(x,j) \bmod p
\]
We have that $(\Gamma,\Pi)$ is useful if every subset of $n$ rows in $A_{\Gamma,\Pi}$ is a linearly independent set of vector over $\Fi^s$. We show later that for any $(\leq n,2)$-expander $\Gamma$, there exists at least one $\Pi$ making $\Gamma$ useful:
\begin{lemma}
\label{lem:useful}
If $\Gamma : [u] \times [t] \to [s]$ is a $(\leq n,2)$-expander, then for $p \geq 2e u$, there exists a $\Pi : [u] \times [t] \to \Fi$ such that $(\Gamma,\Pi)$ is useful.
\end{lemma}

In the cell probe model, we may assume that $\Gamma$ and $\Pi$ are free to evaluate and are known to a data structure since computation is free of charge. With such a pair $(\Gamma,\Pi)$ we may now construct our data structure for $n$-wise independent hashing.

\textit{Construction.}
Initialize the data structure by filling each of the $s$ memory cells by uniformly and independently chosen values in $\Fi$ (the seed). Let $z_0,\dots,z_{s-1}$ denote the values in the memory cells.

\textit{Querying.}
To evaluate $h(x)$ for an $x \in [u]$, compute and return the value
$$
\sum_{j=0}^{t-1} \Pi(x,j)z_{\Gamma(x,j)} \bmod p.
$$

\textit{Analysis.}
Observe that the value returned on the query $x$ equals
$$
\sum_{j=0}^{t-1} \Pi(x,j)z_{\Gamma(x,j)} \bmod p \equiv \sum_{y=0}^{s-1} \sum_{j : \Gamma(x,j)=y} \Pi(x,j)z_{\Gamma(x,j)} \bmod p.
$$
But this is the same as
$(A_{\Gamma,\Pi}z)_x$, i.e. the inner product of the $x$'th row of $A_{\Gamma,\Pi}$ with the randomly drawn vector $z$. Since the rows of $A_{\Gamma,\Pi}$ are $n$-wise independent and $z$ is drawn uniformly, we conclude that the query values $h(0),\dots,h(u-1)$ are $n$-wise independent as well. The query time is $t$ probes and the space usage is $s$ cells of $\lg p$ bits. We thus conclude
\begin{lemma}
\label{lem:hashfromexp}
    Given a bipartite $(\leq n,2)$ expander $\Gamma : [u] \times [t] \to [s]$ and a $p \geq 2e u$, there is a cell probe data structure for evaluating an $n$-wise independent hash function $h : [u] \to \Fi$ using $s$ cells of $w=\Theta(\lg p)$ bits and answering queries in $t$ cell probes.
\end{lemma}
An argument similar to the proof of Lemma~\ref{lem:noncons}, we show the existence of the desired expanders:
\begin{lemma}
\label{lem:nonconshash}
    For any $s \geq 2n$ and any $u \geq n$, there exists a (non-explicit) $(\leq n,2)$ expander $\Gamma : [u] \times [t] \to [s]$ with $t = 2\lg(u/n)/\lg(s/n) + 4$.
\end{lemma}
Combining Lemma~\ref{lem:nonconshash}, Lemma~\ref{lem:useful} and Lemma~\ref{lem:hashfromexp} proves Theorem~\ref{thm:hash}.

What remains is to prove Lemma~\ref{lem:useful} and Lemma~\ref{lem:nonconshash}. We start with Lemma~\ref{lem:useful}.

\begin{proof}(Lemma~\ref{lem:useful})
We give a probabilistic argument. Let $\Gamma : [u] \times [t] \to [s]$ be a $(\leq n,2)$-expander. Draw $\Pi : [u] \times [t] \to \Fi$ by letting $\Pi(x,j)$ be chosen uniformly and independently from $\Fi$. Define an event $E_{\beta}$ for every $\beta \in \Fi^u$ with $1 \leq \|\beta\|_0 \leq n$ ($\|\beta\|_0$ gives the number of non-zeros) that occurs if $\beta A_{\Gamma, \Pi} = 0$. We have that $(\Gamma,\Pi)$ is useful if none of the events $E_{\beta}$ occur. 

Consider one of these events $E_{\beta}$. Since $\Gamma$ is a $(\leq n, 2)$-expander, we have that the set of rows in $A_{\Gamma,\Pi}$ corresponding to non-zero coefficients of $\beta$ have at least $2 \|\beta\|_0$ distinct columns containing an  entry that is chosen uniformly at random and independently from $\Fi$. We thus have $\Pr[E_\beta] \leq p^{-2\|\beta\|_0}$. A union bound finally implies:
\begin{eqnarray*}
\Pr[(\Gamma,\Pi) \textrm{ is not useful}] &\leq& \\
\sum_{i=1}^n \sum_{\beta \in \Fi^u : \|\beta\|_0=i} \Pr[E_\beta] &\leq& \\
\sum_{i=1}^n \binom{u}{i} p^i p^{-2i} &\leq& \\
\sum_{i=1}^n (eu/(ip))^i.
\end{eqnarray*}
For $p \geq 2eu$, this is less than $1$, which concludes the proof of Lemma~\ref{lem:useful}.
\end{proof}

Lastly, we prove Lemma~\ref{lem:nonconshash}.

\begin{proof}(Lemma~\ref{lem:nonconshash}) 
The proof follows that of Lemma~\ref{lem:noncons} uneventfully. Draw $\Gamma$ randomly, with each $\Gamma(x,y)$ uniform and independently chosen in $[s]$. Again, we define an event $E_{S,T}$ for each $S \subseteq [u]$ with $|S| \leq n$ and each $T \subseteq [s]$ with $|T| = 2|S|-1$. The event $E_{S,T}$ occurs if $\Gamma(S) \subseteq T$. We have
\begin{eqnarray*}
    \Pr[\Gamma \textrm{ is not an }(\leq n,2) \textrm{-expander}] &\leq& \\
    \sum_{S,T} \Pr[E_{S,T}] &\leq& \\
    \sum_{i=1}^n \binom{u}{i} \binom{s}{2i} ((2i)/s)^{ti} &\leq& \\
    \sum_{i=1}^n (eu/i)^i(s/(2i))^{2i} ((2i)/s)^{ti} &=& \\
    \sum_{i=1}^n \left(\frac{eu (2i)^{t-3} }{s^{t-2}} \right)^i &\leq& \\
    \sum_{i=1}^n \left(e(u/n) (2n/s)^{t-2} \right)^i 
\end{eqnarray*}
For $s \geq 4n$ and $t \geq 2\lg(u/n)/\lg(s/n) + 4 \geq \lg(u/n)/\lg(s/(2n)) + 4 $, this is less than $1$, completing the proof of Lemma~\ref{lem:nonconshash}.
\end{proof}

\section{Lower Bound for Non-Adaptive Dictionaries}
In this section, we prove cell probe lower bounds for non-adaptive dictionaries supporting membership queries (is $x$ in the input set $X$?). 

We adapt the ``cell-sampling'' technique from~\cite{PTW10}. Roughly speaking, this proof technique shows that there exists a not-too-large subset of cells $C \subseteq [s]$ such that a large number of queries will only
probe cells in $C$ (we say such queries are resolved by $C$) assuming that the query time of the cell probe data structure is impossibly small.
For adaptive and static data structures, it can be observed that the subset of cells $C$ will be different for varying choices of the $n$ input key-value pairs as the probed cells during queries can depend on the memory representation.

For our non-adaptive lower bound, we make the critical observation that the subset of sampled cells $C$ need not depend on the $n$ input key-value pairs. In particular, non-adaptive queries must choose the probed cells without any knowledge of the memory representation. As a result, we are able to separate the adaptive and non-adaptive setting for the dictionary problem and successfully prove a matching lower bound to our constructions as follows:

\begin{proof}(Theorem~\ref{thm:det_lb})
Assume the space usage of a data structure is $s$ cells of $w$ bits each. We assume for the proof that $sw \geq 6 n \lg(u/n)$. For smaller space usage, we can always pad with dummy memory cells. 

For a query $x \in [u]$, let $p(x) \subseteq [s]$ denote the indices of the memory cells probed on query $x$. 

By averaging, for any $q$ with $t \leq q \leq s$, there is a set of $q$ memory cells $C \subseteq [s]$ such that $u \binom{s-t}{q-t}/\binom{s}{q}$ queries $x$ have $p(x) \subseteq C$. Fix such a set $C$. Assume for the sake of contradiction that
$$
t \leq (1/4)\min\left\{q, \frac{\lg(u/n)}{\lg(sw/(n \lg(u/n)))} \right\}.
$$
Then we have
$$
    u \cdot \frac{\binom{s-t}{q-t}}{\binom{s}{q}} =
    u \cdot \frac{q(q-1)\cdots (q-t+1)}{s(s-1)\cdots (s-t+1)} \geq
    u \cdot \left(\frac{(3/4)q}{s}\right)^t.
$$
Letting $q = (1/4)n \lg(u/n)/w$, this is at least $u \cdot \left(\frac{(3/16)n \lg(u/n)}{sw}\right)^t \geq u \cdot \left(\frac{n \lg(u/n)}{sw}\right)^{2t} \geq u \sqrt{n/u} = \sqrt{un}$.

Let $U \subseteq [u]$ denote the set of queries $x$ with $p(x) \subseteq C$. Notice that the memory cells in $C$ serve as a membership data structure for the universe $U$ and inputs $X \subseteq U$ of size $n$. Hence the number of bits in $C$ must be at least $\lg \binom{|U|}{n} \geq (1/2)n \lg(u/n)$. But the cells only have $qw = (1/4)n \lg(u/n)$ bits, a contradiction. We thus conclude:
$$
    t = \Omega\left(\min\left\{ \frac{ n \lg(u/n)}{w}, \frac{\lg(u/n)}{\lg(sw/(n \lg(u/n)))}\right\}\right).
$$
\end{proof}


\section{Conclusion and Open Problems}
\label{sec:concl}
In this work, we presented optimal non-adaptive cell probe dictionaries and data structures for evaluating $n$-wise independent hash functions. 
Our upper bounds rely on the existence of bipartite expanders with quite weak expansion properties, namely $(\leq n,1)$ and $(\leq n,2)$-bipartite expanders. If efficient explicit constructions of such expanders were to be developed, they would immediately allow us to implement our dictionary in the standard word-RAM model. They would also go a long way towards a word-RAM implementation of $n$-wise independent hashing. We thus view our results as strong motivation for further research into such expanders. In personal communication with Bruno Bauwens and Marius Zimand, they have given a preliminary proof that an exciting explicit construction with $s=O(n)$ and $t = (\lg u)^{O(1)}$ exists, thus taking a first step towards an optimal word-RAM implementation.

Next, we remark that our non-explicit constructions of $(\leq n,1)$ and $(\leq n,2)$ expanders are essentially optimal. Concretely, a result of Radhakrishnan and Ta-Shma \cite{RadhakrishnanT00} shows that any $(u,s,t)$-bipartite graph with expansion $1$ requires $t=\Omega(\log (u/n) / \log (s/n))$. 
In more detail, Theorem 1.5 (a) of \cite{RadhakrishnanT00} proves that if $G$ is a $(u,s,t)$-bipartite graph that is an $(n,\epsilon)$ disperser (every set of $n$ left-nodes has at least $(1-\eps)s$ right-nodes), then for $\eps > 1/2$, the left-degree, $t$, is $\Omega(\lg(u/n)/\lg(1/(1-\eps)))$. Since a $(\leq n,1)$-non-contractive expander is also an $(n,\epsilon)$-disperser with $(1-\epsilon) = n/s$, the lower bound $t = \Omega(\log(u/n)/\log (s/n))$ follows.

Finally, we also observe a near-equivalence between non-adaptive data structures for evaluating $n$-wise independent hash functions and non-constructive bipartite expanders. Concretely, assume we have a word-RAM data structure for evaluating an $n$-wise independent hash function from $[u]$ to $[u]$ and assume $w = \lg u$ for simplicity. If the data structure uses $s$ space and answers queries in $t$ time (including memory lookups and computation), then we may obtain an explicit expander from the data structure. Concretely, we form a right node for every memory cell, a left node for every query and an edge corresponding to each cell probed on a query. Now observe that if there was a set of $n$ left nodes $S$ with $|\Gamma(S)| < n$, then from those $|\Gamma(S)|$ memory cells, the data structure has to return $n$ independent and uniform random values in $[u]$. But the cells only have $|\Gamma(S)|w < n \lg u$ bits, i.e. a contradiction. Hence the resulting expander is non-contractive. If the query time of the data structure was $t$, we may obtain the edges incident to a left node simply by running the corresponding query algorithm. Since the query algorithm runs in $t$ time, it clearly  accesses at most $t$ right nodes and computing the nodes to access can also be done in $t$ time. A similar connection was observed by~\cite{ChristianiPT15}.

\bibliographystyle{alpha}
\bibliography{refs}

\newcommand{\etalchar}[1]{$^{#1}$}
\begin{thebibliography}{BMRV02}

\bibitem[AF09]{alon2009power}
Noga Alon and Uriel Feige.
\newblock On the power of two, three and four probes.
\newblock In {\em Proceedings of the twentieth annual ACM-SIAM symposium on
  Discrete algorithms}, pages 346--354. SIAM, 2009.

\bibitem[BBK17]{BBK17}
Joe Boninger, Joshua Brody, and Owen Kephart.
\newblock Non-adaptive data structure bounds for dynamic predecessor search.
\newblock In {\em 37th {IARCS} Annual Conference on Foundations of Software
  Technology and Theoretical Computer Science, {FSTTCS} 2017, December 11-15,
  2017, Kanpur, India}, volume~93 of {\em LIPIcs}, pages 20:1--20:12, 2017.

\bibitem[BHP{\etalchar{+}}06]{Berger2006}
Mette Berger, Esben~Rune Hansen, Rasmus Pagh, Mihai P{\u{a}}tra{\c{s}}cu, Milan
  Ruzic, and Peter Tiedemann.
\newblock Deterministic load balancing and dictionaries in the parallel disk
  model.
\newblock In Phillip~B. Gibbons and Uzi Vishkin, editors, {\em {SPAA} 2006:
  Proceedings of the 18th Annual {ACM} Symposium on Parallelism in Algorithms
  and Architectures, Cambridge, Massachusetts, USA, July 30 - August 2, 2006},
  pages 299--307. {ACM}, 2006.

\bibitem[BL15]{BrodyLarson}
Joshua Brody and Kasper~Green Larsen.
\newblock Adapt or die: Polynomial lower bounds for non-adaptive dynamic data
  structures.
\newblock {\em Theory Comput.}, 11:471--489, 2015.

\bibitem[BMRV02]{Buhrman02}
H.~Buhrman, P.~B. Miltersen, J.~Radhakrishnan, and S.~Venkatesh.
\newblock Are bitvectors optimal?
\newblock {\em SIAM Journal on Computing}, 31(6):1723--1744, 2002.

\bibitem[BZ22]{bauwens2022hall}
Bruno Bauwens and Marius Zimand.
\newblock Hall-type theorems for fast dynamic matching and applications.
\newblock {\em arXiv preprint arXiv:2204.01936}, 2022.

\bibitem[CK09]{CohenKane09}
Jeffrey Cohen and Daniel~M. Kane.
\newblock Bounds on the independence required for cuckoo hashing.
\newblock Manuscript, 2009.

\bibitem[CPT15]{ChristianiPT15}
Tobias Christiani, Rasmus Pagh, and Mikkel Thorup.
\newblock From independence to expansion and back again.
\newblock In Rocco~A. Servedio and Ronitt Rubinfeld, editors, {\em Proceedings
  of the Forty-Seventh Annual {ACM} on Symposium on Theory of Computing, {STOC}
  2015, Portland, OR, USA, June 14-17, 2015}, pages 813--820. {ACM}, 2015.

\bibitem[CW77]{carter1977universal}
J~Lawrence Carter and Mark~N Wegman.
\newblock Universal classes of hash functions.
\newblock In {\em Proceedings of the ninth annual ACM symposium on Theory of
  computing}, pages 106--112, 1977.

\bibitem[DW03]{dietzfelbinger2003almost}
Martin Dietzfelbinger and Philipp Woelfel.
\newblock Almost random graphs with simple hash functions.
\newblock In {\em Proceedings of the thirty-fifth Annual ACM Symposium on
  Theory of Computing}, pages 629--638, 2003.

\bibitem[FKS84]{FredmanKS84}
Michael~L. Fredman, J{\'{a}}nos Koml{\'{o}}s, and Endre Szemer{\'{e}}di.
\newblock Storing a sparse table with 0(1) worst case access time.
\newblock {\em J. {ACM}}, 31(3):538--544, 1984.

\bibitem[GR17]{garg2017set}
Mohit Garg and Jaikumar Radhakrishnan.
\newblock Set membership with non-adaptive bit probes.
\newblock In {\em 34th Symposium on Theoretical Aspects of Computer Science
  (STACS 2017)}. Schloss Dagstuhl-Leibniz-Zentrum fuer Informatik, 2017.

\bibitem[GUV09]{GuruswamiUV09}
Venkatesan Guruswami, Christopher Umans, and Salil~P. Vadhan.
\newblock Unbalanced expanders and randomness extractors from parvaresh-vardy
  codes.
\newblock {\em J. {ACM}}, 56(4):20:1--20:34, 2009.

\bibitem[LPPZ23]{larsen2023optimal}
Kasper~Green Larsen, Rasmus Pagh, Toniann Pitassi, and Or~Zamir.
\newblock Optimal non-adaptive cell probe dictionaries and hashing.
\newblock {\em arXiv preprint arXiv:2308.16042}, 2023.

\bibitem[Pag01]{Pagh01}
Rasmus Pagh.
\newblock On the cell probe complexity of membership and perfect hashing.
\newblock In {\em Proceedings on 33rd Annual {ACM} Symposium on Theory of
  Computing, July 6-8, 2001, Heraklion, Crete, Greece}, pages 425--432. {ACM},
  2001.

\bibitem[PP03]{pagh2003uniform}
Anna Pagh and Rasmus Pagh.
\newblock Uniform hashing in constant time and linear space.
\newblock In {\em Proceedings of the thirty-fifth annual ACM symposium on
  Theory of computing}, pages 622--628, 2003.

\bibitem[PR01]{PaghR01}
Rasmus Pagh and Flemming~Friche Rodler.
\newblock Cuckoo hashing.
\newblock In Friedhelm~Meyer auf~der Heide, editor, {\em Algorithms - {ESA}
  2001, 9th Annual European Symposium, Aarhus, Denmark, August 28-31, 2001,
  Proceedings}, volume 2161 of {\em Lecture Notes in Computer Science}, pages
  121--133. Springer, 2001.

\bibitem[PTW10]{PTW10}
Rina Panigrahy, Kunal Talwar, and Udi Wieder.
\newblock Lower bounds on near neighbor search via metric expansion.
\newblock In {\em 2010 IEEE 51st Annual Symposium on Foundations of Computer
  Science}, pages 805--814. IEEE, 2010.

\bibitem[PY20]{PersianoYao}
Giuseppe Persiano and Kevin Yeo.
\newblock Tight static lower bounds for non-adaptive data structures.
\newblock {\em CoRR}, abs/2001.05053, 2020.

\bibitem[RR18]{RR18}
Sivaramakrishnan~Natarajan Ramamoorthy and Anup Rao.
\newblock Lower bounds on non-adaptive data structures maintaining sets of
  numbers, from sunflowers.
\newblock In {\em 33rd Computational Complexity Conference}, 2018.

\bibitem[RT00]{RadhakrishnanT00}
Jaikumar Radhakrishnan and Amnon Ta{-}Shma.
\newblock Bounds for dispersers, extractors, and depth-two superconcentrators.
\newblock {\em {SIAM} J. Discret. Math.}, 13(1):2--24, 2000.

\bibitem[Sie89]{Siegel89}
Alan Siegel.
\newblock On universal classes of fast high performance hash functions, their
  time-space tradeoff, and their applications (extended abstract).
\newblock In {\em 30th Annual Symposium on Foundations of Computer Science,
  Research Triangle Park, North Carolina, USA, 30 October - 1 November 1989},
  pages 20--25. {IEEE} Computer Society, 1989.

\bibitem[TY79]{tarjan1979storing}
Robert~Endre Tarjan and Andrew Chi-Chih Yao.
\newblock Storing a sparse table.
\newblock {\em Communications of the ACM}, 22(11):606--611, 1979.

\bibitem[Yao81]{Yao81a}
Andrew~Chi{-}Chih Yao.
\newblock Should tables be sorted?
\newblock {\em J. {ACM}}, 28(3):615--628, 1981.

\bibitem[Yeo23]{cryptocuckoo}
Kevin Yeo.
\newblock Cuckoo hashing in cryptography: Optimal parameters, robustness and
  applications.
\newblock In {\em Annual International Cryptology Conference}, pages 197--230.
  Springer, 2023.

\end{thebibliography}

\end{document}